%% file: main.tex
\documentclass[conference]{IEEEtran}
\usepackage{cite}
\usepackage{amsmath,amssymb,amsfonts}
\usepackage{algorithm2e}
\usepackage{graphicx}
\usepackage{textcomp}
\usepackage{xcolor}
\def\BibTeX{{\rm B\kern-.05em{\sc i\kern-.025em b}\kern-.08em
    T\kern-.1667em\lower.7ex\hbox{E}\kern-.125emX}}

\usepackage[hidelinks]{hyperref} % TODO: Is hidelinks allowed for AMC?
\makeatletter
\let\MYcaption\@makecaption
\makeatother
\usepackage{subcaption}

\usepackage[capitalize]{cleveref}
\usepackage{makecell}
\usepackage[all]{hypcap} % TODO: is this allowed? Makes refs jump to the top of the figure instead of to the caption
\usepackage{siunitx} % TODO: debatable if we need this only for percentage values, but it DOES change the spacing in that case
\usepackage[nonumberlist]{glossaries}
\usepackage{glossary-inline}

\loadglsentries[main]{glosfile}
\glsdisablehyper % Disables acronym hyperlinks (they only work with a glossary, otherwise they just point to the beginning of the paper)

\usepackage[colorinlistoftodos,prependcaption,textsize=tiny]{todonotes} % TODO Remove

\allowdisplaybreaks

\newcommand{\tool}{MILQ}

\makeatletter
\newcommand{\linebreakand}{%
  \end{@IEEEauthorhalign}
  \hfill\mbox{}\par
  \mbox{}\hfill\begin{@IEEEauthorhalign}
}
\makeatother

\newcommand\copyrighttext{%
  \footnotesize  © 20xx IEEE. Personal use of this material is permitted. Permission from IEEE must be
obtained for all other uses, in any current or future media, including
reprinting/republishing this material for advertising or promotional purposes, creating new
collective works, for resale or redistribution to servers or lists, or reuse of any copyrighted
component of this work in other works.}
\newcommand\copyrightnotice{%
\begin{tikzpicture}[remember picture,overlay]
\node[anchor=south,yshift=10pt] at (current page.south) {\fbox{\parbox{\dimexpr\textwidth-\fboxsep-\fboxrule\relax}{\copyrighttext}}};
\end{tikzpicture}%
}

\begin{document}

\title{Multithreaded parallelism for heterogeneous clusters of QPUs  \\
}

\author{
\IEEEauthorblockN{Philipp Seitz}
\IEEEauthorblockA{\emph{Technical University of Munich}\\
\emph{TUM School of Computation, Information and Technology}\\
\emph{Department of Computer Science}\\
Boltzmannstra{\ss}e 3, 85748 Garching, Germany}
\IEEEauthorblockA{\href{https://orcid.org/0000-0003-3856-4090}{0000-0003-3856-4090}}
\and
\IEEEauthorblockN{Manuel Geiger}
\IEEEauthorblockA{\emph{Technical University of Munich}\\
\emph{TUM School of Computation, Information and Technology}\\
\emph{Department of Computer Science}\\
Boltzmannstra{\ss}e 3, 85748 Garching, Germany}
\IEEEauthorblockA{\href{https://orcid.org/0000-0003-3514-8657}{0000-0003-3514-8657}}
%\and
\linebreakand
\IEEEauthorblockN{Christian~B.~Mendl}
\IEEEauthorblockA{\emph{Technical University of Munich}\\
\emph{TUM School of Computation, Information and Technology} \\
\emph{Department of Computer Science} and \emph{TUM Institute for Advanced Study}\\
Boltzmannstra{\ss}e 3, 85748 Garching, Germany}
\IEEEauthorblockA{\href{https://orcid.org/0000-0002-6386-0230}{0000-0002-6386-0230}}
% \and
% \IEEEauthorblockA{\emph{dept. name of organization (of Aff.)} \\
% \emph{name of organization (of Aff.)}\\
% \and
% \IEEEauthorblockA{\emph{dept. name of organization (of Aff.)} \\
% \emph{name of organization (of Aff.)}\\
}

\maketitle

\begin{abstract}
% What the paper is about:
In this work, we present \tool{}, a quantum unrelated parallel machines scheduler and cutter.
The setting of unrelated parallel machines considers independent hardware backends, each distinguished by differing setup and processing times.
\tool{} optimizes the total execution time of a batch of circuits scheduled on multiple quantum devices.
It leverages state-of-the-art circuit-cutting techniques to fit circuits onto the devices and schedules them based on a mixed-integer linear program. 
Our results show a total improvement of up to {\boldmath$\SI{26}{\percent}$} compared to a baseline approach. 
\glsresetall
\end{abstract}
% *************************

\copyrightnotice{}
\section{Introduction}\label{sec:Introduction}

Quantum computing promises exponential speedup compared to classical computing for certain computational tasks~\cite{shor1994algorithms}.
To realize such a quantum advantage with fault tolerance, the estimated number of required qubits is roughly $10^8$\cite{fowler2012}, considerably more than the number provided by currently available devices (for example, IBM Osprey with $433$ qubits).
% MG: I saw two papers that didn't cite Osprey (e.g., one of them was "QFactor: A Domain-Specific Optimizer for Quantum Circuit Instantiation")

This is due to overhead introduced by error correction, which is necessary because of the error-prone hardware realizations of qubits.
In the near term, only small, noisy quantum devices are available, and their access is limited.
Still, many supercomputing centers have already started integrating quantum devices on their premises.
Multiple physical realizations of quantum computers exist (called modalities), such as superconducting qubits~\cite{kjaergaard2020}, neutral atoms~\cite{Wu2021}, or ion traps~\cite{Bruzewicz2019}, each with unique characteristics.
To diversify their portfolio, supercomputing centers prepare heterogeneous infrastructures supporting multiple modalities.
In this article, we propose a multithreading scheme for such a platform to maximize utilization.
We also implement a prototype, available as a repository on GitHub at \url{https://github.com/qc-tum/milq}.

% The article is structured as follows.
% In \Cref{sec:Motivation}, we describe a use-case that motivates our development.
% \Cref{sec:BackgroundAndRelatedWork} covers the basics of quantum computing and introduces related subjects.
% We present our software model in \Cref{sec:Proposal}.
% The underlying model is described in \Cref{sec:ProblemStatement} followed by the evaluation in \Cref{sec:Results}.
% We finally discuss future work in \Cref{sec:FutureWork} and conclude in \Cref{sec:Conclusion}.

\section{Motivation}\label{sec:Motivation}

In the \gls{NISQ} era, quantum resources are still scarce.
While the individual device sizes keep growing, access remains limited.
Unfortunately, most circuits do not fit perfectly on a device.
If the circuit is too small, this leads to underutilization.
If the circuit is too large, it can be cut into fitting parts, but likely with a remainder that underutilizes the \gls{QPU}.
This is especially wasteful, considering that devices are exclusively assigned to users.
As an extreme example, running a variational algorithm can block a device for a long time, even if enough qubits are available to run other simulations.
\tool{} solves this issue by considering all submitted circuits before running them.
It resizes circuits appropriately and uses available resources in parallel.
As a result, it is possible to have multiple circuit instances running on one \gls{QPU}, hence the term ``multithreading''. 

Due to its modular nature, \tool{} extends beyond \gls{NISQ}.
One can integrate communication overhead as an additional constraint when considering distributed quantum systems.
It is also relevant as a case study since many techniques apply to near-term integration scenarios.
Many supercomputing centers have started adopting \glspl{QPU} as novel accelerators.
In this domain, scheduling is a common problem.
\tool{} is an initial attempt at providing a runtime component that can be expanded further.

\section{Background and Related Work}\label{sec:BackgroundAndRelatedWork}

\tool{} combines techniques from various application areas.
In this section, we provide the necessary background knowledge and investigate comparable solutions.
The scheduling component is based on a \gls{MILP}, hence the name \tool{}.

The scheduling of quantum circuits is still an emerging problem.
No established tool exists; most existing schedulers are based on simple first-in, first-out concepts.
One can consider scheduling as a part of the mapping problem if one includes the mapping over distributed devices~\cite{bandic2023}.
A quantum circuit can be arbitrarily distributed, assuming gate teleportation between devices is possible, which is not necessarily true.
Bhoumik et al.~\cite{bhoumik2023} consider the scheduling problem regarding error mitigation by finding optimal mappings from (cut) subcircuits to multiple \glspl{QPU}.
Their work is based on an integer linear program, which optimizes for circuit fidelity.

\subsection{Scheduling}\label{ssec:Scheduling}
General scheduling is an optimization problem studied in operations research and computer science.
Using the Graham $\alpha|\beta|\gamma$ classification scheme~\cite{graham1979}, the problem of interest in this article can be described as $R|\text{res}_1|C_{\max}$.
It is a variant of the unrelated parallel machine problem, which has been studied extensively in literature~\cite{fanjul2011,rodriguez2013,alharkan.2019,Lenstra1990}.
$R$ indicates $m$ completely independent machines (in our case the \glspl{QPU}), which use a single shared resource $\text{res}_1$, namely qubits.
The problem is optimized for the makespan $C_{\max}$, the latest completion time of any job.
%\discuss{MG}{The more specific problem of related parallel machines is NP-hard~\cite{Karp1972}.} % MG: Our problem is a variant of the unrelated parallel machine problem, so why is it mentioning the related parallel machines problem? Also, which of those is meant by "the problem" in the next sentence?
Unfortunately, it is NP-hard to solve this exactly~\cite{Lenstra1990}. Hence, such problems are usually stated as a \gls{MILP} and solved using heuristic approaches.
Given the resource constraints, the goal is to provide a schedule where each job is assigned to one machine, and the overall execution time is minimized.
Standard techniques involve Simulated Annealing, Tabu Search, and genetic algorithms~\cite{glass1994}.
Depending on the availability of the job information, algorithms either work offline when all information is known ahead of time or online when information is only accessible right before scheduling.
A variant of the online version considers batches of jobs that are scheduled at the same time.
 
In computer science, scheduling is relevant in operating systems and \gls{HPC}.
Users of an \gls{HPC} cluster submit jobs to the system, which are initially queued.
Then, they are assigned a portion of the available resources for a fixed amount of time.
Typically, the exact resource assignment is not disclosed to the user.
The requests are usually handled by a resource management tool, most commonly SLURM~\cite{Yoo2003}, which also provides diagnostics and monitoring utilities.
System utilization and overall runtime are the relevant metrics. 

\subsection{High Performance Computing and Quantum Computing}\label{ssec:HPCQC}

Integrating \gls{QC} into \gls{HPC} is an ongoing process.
Emerging microarchitectures and multiple integration scenarios are part of the challenges in this domain~\cite{humble2021}.
\gls{QC} is transitioning out of the laboratories, and computing centers are starting to integrate quantum hardware into their premises.
There is no established straightforward blueprint for combining classical with quantum infrastructure.
In this state of uncertainty, the computing centers provide multiple modalities with varying capabilities.
Emerging standards for common interfaces, like QIR~\cite{QIRSpec2021}, abstract the hardware details from the user.
Still, most algorithms are run on a single device based on hardware availability.
This leads to underutilization, as the entire device is reserved for the whole duration of the algorithm.
As a user, this is also frustrating; the time-to-solution increases due to the longer queueing times.

\subsection{Circuit Knitting}\label{ssec:CircuitKnitting}

In the \gls{NISQ} era, hardware is physically limited, restricting the possible circuits in two ways.
First, the \emph{width} of a circuit is restricted by the number of available qubits; second, the \emph{depth} of a circuit is limited by the decoherence times of the qubits.
\emph{Circuit Knitting} (also called \emph{Circuit Cutting}) is a method to resize circuits in both dimensions.
This comes at the cost of additional sampling overhead and classical computation.

\begin{figure}
    \centering
    \includegraphics[scale=1]{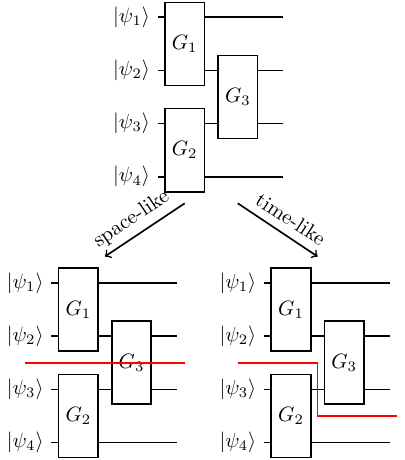}
    \caption{Circuit knitting in two variants: space-like and time-like cuts are possible.}
    \label{fig:cutting}
\end{figure}

\emph{Time-like} cuts reduce circuit depth by cutting wires in a circuit and executing the resulting partial circuits at different times~\cite{peng2020}.
\emph{Space-like} cuts, on the other hand, reduce circuit width by decomposing (multi-qubit) gates~\cite{mitarai2021}.
Both techniques reconstruct the expectation value of the original circuit by sampling from a quasi-probability distribution from the resulting sub-circuits.
The procedures are depicted in \cref{fig:cutting} in a simplified manner.
Recent works~\cite{Ufrecht2023,harada2023optimal} keep improving sampling overhead with advanced techniques. 

\section{Proposal}\label{sec:Proposal}

\begin{figure}
    \centering
    \includegraphics[scale=.5]{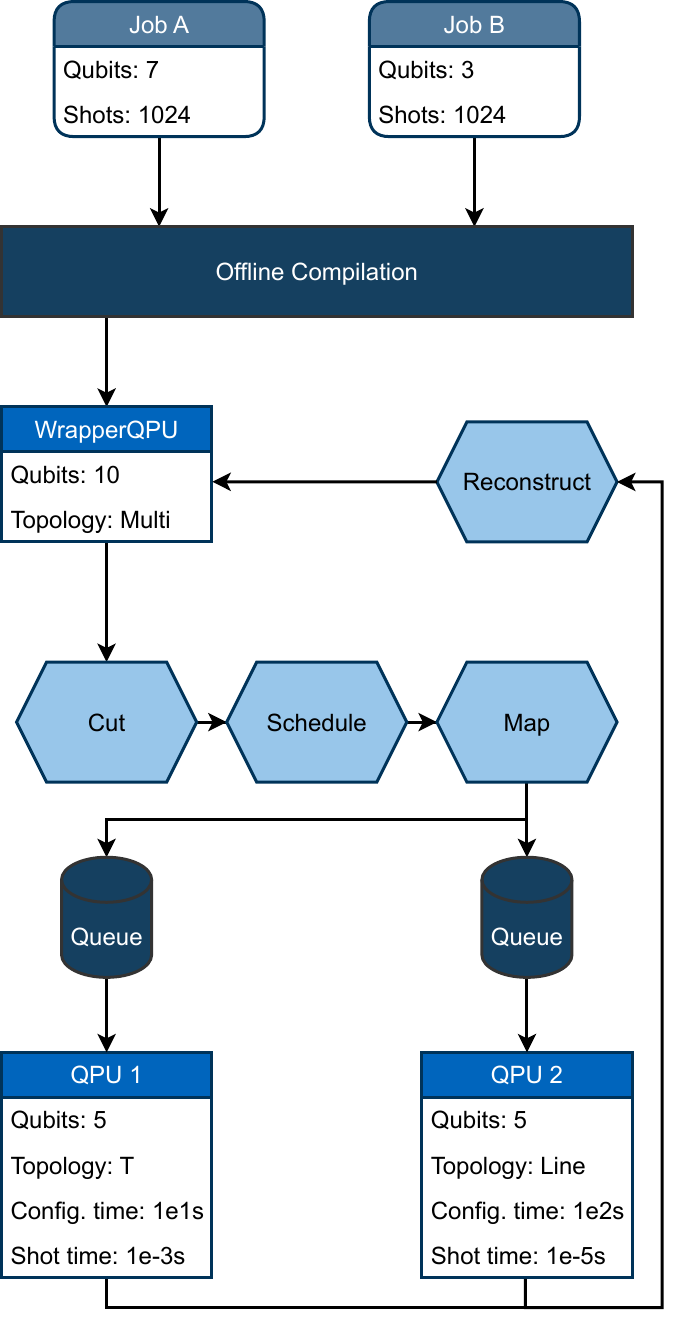}
    \caption{Simplified overview of the intended workflow. }
    \label{fig:multithreading}
\end{figure}

\tool{} is a standalone project built on the infrastructure of the popular quantum software framework Qiskit~\cite{Qiskit}.
The intended workflow is summarized in \cref{fig:multithreading}.
Users build their circuits with Qiskit and specify multiple hardware backends. 
For example, two jobs \emph{A} and \emph{B} are submitted to a system that combines the two fictional devices \emph{QPU 1} and \emph{QPU 2}.
The circuits are compiled individually before being submitted to the system; any compilation is treated as a black box.
% MG: This terms is only defined later
After compilation, the circuits are submitted as a job to a joint interface.
The system greedily resizes circuits using knitting techniques to fit the available hardware and keeps track of execution for reconstruction during postprocessing.
The cutting component supports only gate-cutting for the moment.
Only necessary cuts are selected based on the size of the \glspl{QPU}; the optimality conditions to reduce the sampling overhead are not considered yet. % TODO we can at least add permutations
For scheduling, \tool{} solves a linear programming task, which we describe in \Cref{sec:ProblemStatement} in more detail.
The goal of the optimization is to minimize the overall execution time.
Once the hardware is determined, a hardware-specific compilation procedure prepares the circuits for execution.
After obtaining the results, \tool{} reconstructs the measurement data and assigns it to the correct circuits.

\tool{} comprises three main components: a \emph{QPU wrapper}, a \emph{scheduler}, and a \emph{compilation pipeline}.
The \emph{QPU wrapper} is an abstraction layer, imitating the behavior of a single \gls{QPU}.
This minimizes the necessary code changes when switching hardware providers.
Still, \tool{} can use the available hardware information when producing a schedule, and hardware of different modalities is supported.

We assume that there exists a modular \emph{compilation pipeline}.
Modularity provides the possibility of running compilation steps at different times.
While we treat most of the compilation as a black box, \tool{} assumes that compilation can be split up over several phases:
\emph{offline compilation}, which is hardware agnostic, and \emph{online compilation}, which can use hardware information.
This mimics the workflow of an \gls{HPC} application, where jobs are fully compiled before submission.
The knitting module in \tool{} (based on~\cite{circuit-knitting-toolbox}) requires $n$-qubit gates to be decomposed first.
This can be achieved offline, using synthesis tools or the built-in functionality from Qiskit~\cite{Qiskit}. %  such as BSQKit~\cite{bqskit}
Hiding implementation details is one benefit of the abstraction layer.
For example, hardware-specific optimizations and necessary modifications, such as mapping, are abstracted.

During the scheduling process, \tool{} analyzes a batch of circuits and provides an optimal schedule based on a \gls{MILP} (see \Cref{sec:ProblemStatement}).
The scheduler assumes that circuits already fit the available hardware.
The system queries the backends for each circuit's estimated processing and setup times.
We use dummy data based on the circuit depth as accurate data from the hardware interface is unavailable.
The scheduler supports two modes of operation: \emph{simplified} and \emph{extended}.
In the simplified variant, we assume that the setup times depend only on the current job independent of its predecessor, which relaxes the constraints resulting from \Cref{eq:successor}.

\section{Problem Statement}\label{sec:ProblemStatement}

We formulate the scheduling of circuits as an (offline) \gls{MILP}.
The problem is a more constrained variant of the well-studied unrelated parallel machines problem discussed in~\Cref{ssec:Scheduling}.
Each job (quantum circuit) can be assigned to one machine (\gls{QPU}) in this model. 
Depending on this machine, each job has a unique processing time.
Each job has a machine- and sequence-dependent setup time.
This mimics, for example, the reconfiguration of the \glspl{AWG} between experiments. 
To model qubit numbers, each machine has a fixed capacity, which cannot be exceeded at any time.
The notation is summarized in \Cref{tab:notation}.
$\mathbb{M}$ and $T_{\max}$ have to be tuned following the magnitude of the input parameters $p_{im}$ and $s_{ijm}$.
One key difference compared to existing models is the relaxation of the succession constraints.
Typically, each job has one predecessor and up to one successor.
In our scenario, however, multiple jobs can run in parallel on one machine, meaning one job can have multiple successors. Hence, we use the following definition for the successor relation (cf.~\Cref{tab:notation}):
\begin{equation} \label{eq:successor}
    y_{ijm} = 1 \Longleftrightarrow c_i < c_j \land \nexists k \in J: c_i < c_k < b_j \land \gamma_{ijm} \land \gamma_{i k m}
\end{equation}
Paraphrased, a job $j$ is the successor of job $i$ on machine $m$ when no other job $k$ was completed in between.
To allow for more than one successor, we also relax the completion time constraint \eqref{eq:con5} such that having multiple predecessors does not add a penalty by having to set up twice.
With this, we implicitly assume that the setup time for multiple circuits can be combined.
Otherwise, mapping all possible combinations of circuits to the \gls{MILP} would exponentially increase the complexity of the problem.

\begin{table}
    \caption{MILP Notation}
    \label{tab:notation}
    \begin{tabular}{l|r}
        \hline
        \multicolumn{2}{|c|}{Metavariables} \\ \hline
        $\mathbb{M}$ & A big number \\ 
        $T_{\max}$ & A large number of time slots \\ % TIMESTEPS
        $0$ & Dummy job \\ \hline
        \multicolumn{2}{|c|}{Input Parameters} \\ \hline
        $J$ & Set of jobs (circuits) \\
        $M$ & Set of machines (QPUs) \\
        $p_{im}$ & Processing time of job $i$ on machine $m$ \\ %  (\emph{p\_times})
        $s_{ijm}$ & Setup time of job $j$ after job $i$ on machine $m$ \\ %  (\emph{s\_times})
        $q_{i}$ & Required resources (qubits) of job $i$  \\ % (\emph{job\_capacities})
        $Q_{m}$ & Available resources (qubits) of machine $m$  \\ \hline % (\emph{machine\_capacities})
        \multicolumn{2}{|c|}{Indices} \\ \hline
        $i,j,k$ & Index of jobs, $i,j,k \in J$ \\
        $m$   & Index of machines, $m \in M$ \\
        $t$   & Timesteps, $t \in T = \{0,1,2,\dots,\mathrm{T_{\max}}\}$ \\ \hline
        \multicolumn{2}{|c|}{Binary Decision Variables} \\ \hline
        $x_{im}$ & 1 if job $i$ is scheduled on machine $m$\\
        $y_{ijm}$ & 1 if job $j$ is is a successor of job $i$ on machine $m$ \\
        $z_{imt}$ & 1 if job $i$ is scheduled on machine $m$ at timestep $t$ \\ \hline
        \multicolumn{2}{|c|}{Binary Helper Variables} \\ \hline
        $\alpha_{ij}$ & 1 if job $i$ completes before job $j$ starts\\
        $\beta_{ij}$ & 1 if job $i$ completes before job $j$ completes \\
        $\gamma_{ijm}$ & 1 if jobs $i,j$ are scheduled on machine $m$\\
        $\delta_{ijkm}$ & \makecell[r]{1 if job $k$ completes after $i$ completes \\ but before $j$ starts on $m$} \\ \hline
        \multicolumn{2}{|c|}{Real Decision Variables} \\ \hline
        $c_j$ & Completion time of job $j$ \\
        $c_{\max}$ & Makespan \\
        $b_j$ & Start time of job $j$ \\ \hline
    \end{tabular}
\end{table}

\subsection{MILP Formulation}

The notation in ~\Cref{tab:notation} and problem formulation are derived from Al-harkan and Qamhan~\cite{alharkan.2019}.
The optimization problem is formulated as follows:

\newcommand{\constraint}[1]{\tag{C#1}\label{eq:con#1}}
\begin{align}
    & \min(c_{\max}) \tag{OBJ}\label{eq:objective}\\  % Makespan minimization \\
    & \text{Subject to:}\nonumber\\
    & c_j \leq c_{\max} \quad \forall j \in J \constraint{1}\\ % Makespan constraint $j$ \\
    & c_0 = 0 \constraint{2}\\ % Completion time of dummy job $0$ \\
    & \sum_{m\in M} x_{jm} = 1 \quad \forall j \in J \constraint{3}\\ % Each job is scheduled on exactly one machine \nonumber \\
    & \sum_{m\in M} z_{jmt} \leq 1 \quad \forall j \in J, \forall t \in T\constraint{4} \\ % Each job runs on at most one machine at any timestep \\
    & c_j \geq b_j + \sum_{m \in M}p_{jm}\cdot x_{jm} + \sum_{i\in J\cup\{0\}}\sum_{m\in M}s_{ijm}\cdot y_{ijm} \nonumber\\  &  \forall j \in J \constraint{5}\\ % Completion time of job $j$ \\
    & b_j \geq c_i + \mathbb{M} \cdot  \left(\sum_{m\in M}y_{ijm} - 1\right) \nonumber\\  & \forall j \in J, \forall i \in J\cup\{0\} \constraint{6}\\ % Start time of job $j$ after predecessor $i$ has completed? \\
    &\sum_{m\in M}\sum_{t\in T} z_{jmt} = c_j - b_j + 1 \quad \forall j \in J \constraint{7}\\ % Each job is fully scheduled \\
    & \sum_{t\in T} z_{jmt} \leq \mathbb{M} \cdot  x_{jm} \quad \forall j \in J, \forall m \in M \constraint{8}\\ % Each job is scheduled on a machine for its entire duration \\
    & c_j \geq t\cdot  \sum_{m\in M} z_{jmt} \quad \forall j \in J, \forall t \in T \constraint{9}\\ % Completion time of job $j$ \\
    & s_j \leq t\cdot  \sum_{m\in M} z_{jmt} + \mathbb{M} \cdot  \left(1 - \sum_{m\in M} z_{jmt}\right) \nonumber\\  & \forall j \in J, \forall t \in T \constraint{10}\\ % Completion time of job $j$ \\
    & \sum_{j\in J} q_{j} \cdot  z_{jmt} \leq Q_{m} \quad \forall m \in M, \forall t \in T\constraint{11} \\ % Each machine is not overloaded 
    & 1 \leq \sum_{i\in J\cup\{0\}}\sum_{m\in M} y_{ijm} \quad \forall j \in J \constraint{12}\\ 
    & \mathbb{M} \cdot  x_{jm} \geq  \sum_{i\in J\cup\{0\}} y_{ijm} \quad \forall j \in J, \forall m \in M \constraint{13}\\
    & \mathbb{M} \cdot  x_{jm} \geq  \sum_{i\in J\cup\{0\}} y_{jim} \quad \forall j \in J, \forall m \in M \constraint{14}\\
    & z_{jm0} = y_{0jm} \quad \forall j \in J \forall k \in M \constraint{15}\\
    & \mathbb{M} \cdot  \alpha_{ij} \geq b_j - c_i \quad \forall i,j \in J, i\neq j \constraint{16} \\
    & \mathbb{M} \cdot  \beta_{ij} \geq c_j - c_i \quad \forall i,j \in J, i\neq j \constraint{17} \\
    & \gamma_{ijm} \geq x_{im} + x_{jm} - 1  \quad \forall i,j \in J, i\neq j, \forall m \in M \constraint{18}\\ 
    & \delta_{ijkm} \geq \alpha_{kj} + \beta_{ij} + \gamma_{ijm} + \gamma_{ikm} - 3 \nonumber\\ & \forall i,j,k \in J, i\neq j, \forall m \in M \constraint{19} \\ 
    & y_{ijm} \geq \alpha_{ij} + (1 - \sum_{k \in J} \delta_{ijkm}) + \gamma_{ijm} - 2 \nonumber\\ & \forall i,j \in J, \forall m \in M \constraint{20}
\end{align}

The overall goal, formulated in the objective function \eqref{eq:objective}, is to minimize the makespan. 
The constraints \eqref{eq:con1} bound it on the maximal completion time of any job.
Constraints \eqref{eq:con3} and \eqref{eq:con4} ensure that a job is executed on one machine at a time.
\eqref{eq:con5} calculate the completion time, and \eqref{eq:con6} ensures that all predecessors of a job are done before starting it.
Constraints \eqref{eq:con7} ensure entire processing of a job; \eqref{eq:con8} align the time-dependent execution variables such that they match the chosen machine.
\eqref{eq:con9} and \eqref{eq:con10} fix the execution between start end completion.
\eqref{eq:con11} constrain the resource usage.
The previous constraints could easily be adapted to time-dependent resource availability to model erroneous qubits or partial recalibration procedures.
The constraints \eqref{eq:con12}--\eqref{eq:con19} ensure the validity of the successor relationship.
Each regular job has a predecessor \eqref{eq:con12}, which is on the same machine (\eqref{eq:con13} and \eqref{eq:con14}).
Jobs running at $t=0$ are constrained to use the dummy job with \eqref{eq:con15}.
\eqref{eq:con16}--\eqref{eq:con19} set up the helper variables, and the successor is finally set in the constraint \eqref{eq:con20}.

\subsection{Example Assignment} \label{ssec:example}
As a simple example, we look at integer processing and setup times.
This drastically reduces the number of necessary time-step variables.
In this example, we revisit the setting in \cref{fig:multithreading} for the circuits and \glspl{QPU}.
Initially, circuit \emph{A} does not fit on any available device.
By performing a greedy cut on \emph{A}, \tool{} generates four five-qubit jobs and four two-qubit jobs. 
Together with the three-qubit job \emph{B}, these jobs are subsequently distributed across two five-qubit devices.

% \Cref{tab:capacity} shows all resulting jobs after cutting the circuit.

% \begin{table}
%     \centering
%     \caption{Job and machine specific capacities ($q_i, Q_k$).}
%     \label{tab:capacity}
%     \begin{tabular}{cc}
%         \begin{tabular}[t]{cc}
%     \toprule
%         Job & Capacity \\ \midrule
%         0 & 0 \\
%         A & 5 \\
%         B & 5 \\
%         C & 5 \\
%         D & 5 \\
%         E & 3 \\
%         F & 2 \\
%         G & 2 \\
%         H & 2 \\
%         I & 2 \\
%     \bottomrule
%     \end{tabular} &
%         \begin{tabular}[t]{lc}
%     \toprule
%         Machine & Capacity \\ \midrule
%         QPU 1 & 5 \\
%         QPU 2 & 5 \\
%     \bottomrule
%     \end{tabular}
%     \end{tabular}
% \end{table}

We generate synthetic processing times by randomly applying uniformly sampled variations to the circuit size.  %listed in \Cref{tab:processing_time}.
In reality, the difference in execution time between single-circuit executions is negligible.
Still, this can accumulate significantly when considering the high number of shots in \gls{NISQ} experiments.
Setup times are generated pairwise, depending on the size of both circuits.
% \discuss{MG}{Similarly}, we incorporate two realistic scenario properties when synthesizing the sequence-specific setup data for the extended algorithm.
% % MG: What does this Similarly refer to? For me, it implies that we also incorporated realistic scenario properties when synthesizing processing times
% First, sampling overhead is introduced by cutting techniques.
% \discuss{MG}{Second, execution times between modalities differ; neutral atom systems are roughly two magnitudes slower than superconducting systems.} % MG: But how is this modeled by our sequence-specific setup times??
% \discuss{MG}{One caveat is the simplified binary relation.
In reality, \tool{} could query this information during scheduling if the backends can estimate this accurately.
Also, the setup time would depend on all combinations between predecessors and successors.
For this example, we use the maximum of the possible setup times. 

\begin{figure}
    \centering
    \includegraphics[scale=0.5]{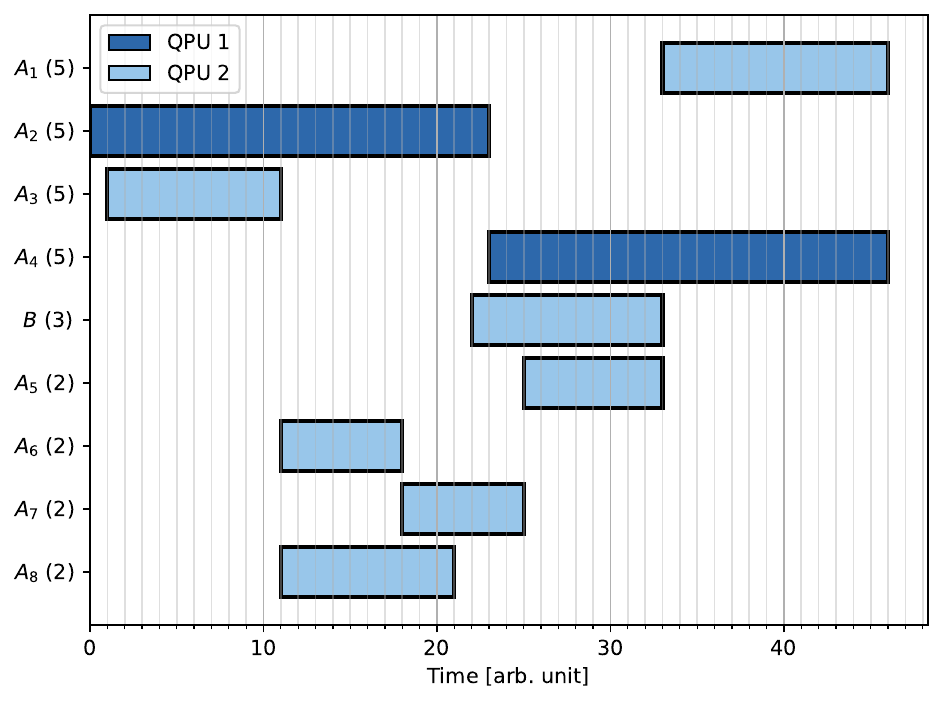}
    \caption{Schedule of the sample problem obtained by the \emph{simple} schedule.}
    \label{fig:example}
\end{figure}

The resulting \gls{MILP} comprises $3188$ ($1208$) variables and $3272$ ($1355$) constraints for the \emph{extended} (\emph{simple}) algorithm.
We run the example using the Gurobi~\cite{gurobi} solver on a single 80-way Intel Ice Lake node.
The resulting \emph{simple} schedule is depicted in \cref{fig:example} and takes approximately $4$~minutes to be generated.
Due to resource limitations, we stop the \emph{extended} model when the gap between the upper bound and the incumbent falls below \SI{20}{\percent}, which still takes $8.6$ hours.
As parameters, we choose $\mathbb{M}=1000$ and $T_{\max}=64$.

\section{Results}\label{sec:Results}

To validate \tool{}, we benchmark multiple batches of random circuits.
We compare the resulting schedule of \tool{} with a baseline implementation of a simplified algorithm.
The system under test is the \emph{scheduling} component of \tool{}.
Every other component is fixed, especially the circuit knitting and mapping components.
Our primary focus lies on the overall makespan, but we also provide insight into the real-time performance of the algorithms.

\subsection{Hardware Analysis}

Unfortunately, setup and processing times are often not directly available from vendors, even less so preliminary estimates of those numbers for a given job. Hence, we fall back to using dummy values.
We argue, however, that modeling setup and processing times as machine- and job-specific and, in the first case, additionally as sequence-dependent values is a reasonable choice.
On the one hand, gate times between different hardware implementations can vary by orders of magnitude.
For instance, gate times on superconducting platforms can range from $10$~nanoseconds up to microseconds~\cite{kjaergaard2020}.
The processing time of a circuit mainly depends on its depth and the number of shots.
On the other hand, we justify sequence-dependent setup times by platform-specific hardware configurations.
For instance, platforms utilizing neutral atoms can reconfigure the layout of atom traps as part of their experimental setup. 
This reconfiguration allows for three-dimensional topologies and scales according to the number of layers in the mapping~\cite{Barredo2018}. 
Although current commercial compilers do not yet support such functionalities, it is a plausible feature for the future, making it an ideal scenario for testing \tool{}.

% We analyzed existing hardware regarding their processing and setup times as the basis for our considerations.
% Unfortunately, vendors rarely disclose this information.
% For the IBM device IBM~Lagos, we analyze the transpilation time and the execution time. 
% Both scale with the depth of the circuit; hence, this validates our approach in using job and machine-dependent parameters for setup- and processing times.
% We justify using sequence-dependent setup times with hardware configuration. For example, neutral atom platforms allow reconfiguration of the atom trap layout as part of their experiment setup.
% This can be done for three-dimensional topologies and scales with the number of layers in the mapping~\cite{Barredo2018}.
% Commercial compilers do not support this yet, but it is a likely feature in the future, \add{PSe}{which would make this an ideal test case.}

\subsection{Benchmarks}

As a benchmark, we look at randomly selected circuits of various sizes using the MQT Bench~\cite{quetschlich2023mqtbench} tool.
% MG: Qiskit is not emphasized at all (like Gurobi). MQT Bench is emphasized only once. Other stuff ("QPU wrapper") is emphasized all the time. -> Consistency needed
We generate ten batches of seven circuits with the maximum size restricted to the largest available device.
This mimics the situation after resizing the circuits.
From the MQT Bench, we select the ``random'' option with optimization level zero.
We compare two scenarios, one with the same configuration as in \Cref{ssec:example} (two \glspl{QPU} with capacity $5$) and the second scenario with three \glspl{QPU} with the sizes ($5$, $6$, $20$).
For both scenarios, the two models \emph{simple} and \emph{extended} are evaluated against a \emph{baseline} algorithm (see \Cref{ssec:Baseline}).
We randomly generate processing and sequence-dependent setup times in both scenarios based on the hardware estimations for superconducting platforms.
Additionally, we run a set of trials with real-valued times and an independent set of integer-valued times.
The simplified model assumes only machine-dependent setup times, which we generate by taking the maximum $s_{jm} = \max_{i\in J} s_{ijm} \forall j \in J$.
After generating the schedules, we recalculate the makespan based on the original setup times, using the successor relation from \Cref{eq:successor}.

\subsection{Baseline}\label{ssec:Baseline}

As a baseline algorithm, we use an adapted version of first-fit decreasing bin packing~\cite{baker1985}.
This uses the simplified assumption that processing and setup times are independent of jobs and machines and are equally long.
Each bin represents a \gls{QPU}, and copies thereof model the simplified succession of jobs.
\Cref{alg:ffd} summarizes the procedure; a copy of all instances is added instead of opening a single bin to ensure equal distribution over all devices.
This provides a naive approach while still allowing multithreading.
Compared to sequential schedules for each device, this is already an improvement.
% \begin{algorithm}
% \SetAlgoLined
% \KwData{$J=: \text{set of jobs}, B:= \text{set of bins}$}
% \KwResult{Assignment of jobs to bins}
% $j \gets \text{list}(\text{sort}(J,decreasing))$\;
% $bins \gets [\ ]$\;
% \While{$j \neq \emptyset$}{
%     $bins.\text{append}(\text{list}(B))$\;
%     \For{$job \in j$}{
%         \If{$bins.\text{fits}(job)$}{
%             $bins.\text{insert}(job)$\;
%             $j.\text{remove}(job)$\;
%         }
%         \lElse{\textbf{break}}
%     }
% }
% \textbf{return} $bins$\;
% \caption{Scheduling with first-fit decreasing bin packing.}\label{alg:ffd}
% \end{algorithm}
% \MG{My version of the algorithm:}
\begin{algorithm}[t!]
\SetKwInOut{Input}{Input}
\SetKwInOut{Output}{Output}
\SetKwFunction{Sorted}{sorted}
\SetKwFunction{Find}{FindFirstFitting}

\Input{$J:= \text{set of jobs}, B:= \text{list of bins}$}
\Output{Bins filled with jobs}
$J' \gets \Sorted{J, qubit\_count descending}$\;
$open \gets B$\;
$closed \gets \emptyset$\;
\ForEach{$job \in J'$}{
    $bin \gets \Find{job, open}$\;
    \If{$bin$ is none}{
        $new \gets B$\;
        $bin \gets \Find{job, new}$\;
        extend $open$ with $new$\;
    }
    add $job$ to $bin$\;
    \If{$bin$ is full}{
        remove $bin$ from $open$\;
        add $bin$ to $closed$\;
    }
}
add all $open$ to $closed$\;
\KwRet{$closed$}\;
\caption{Scheduling with first-fit decreasing bin packing.}\label{alg:ffd}
\end{algorithm}

\begin{figure*}[!ht]
    \centering
    \begin{subfigure}{0.45\textwidth}
            \includegraphics[scale=0.5]{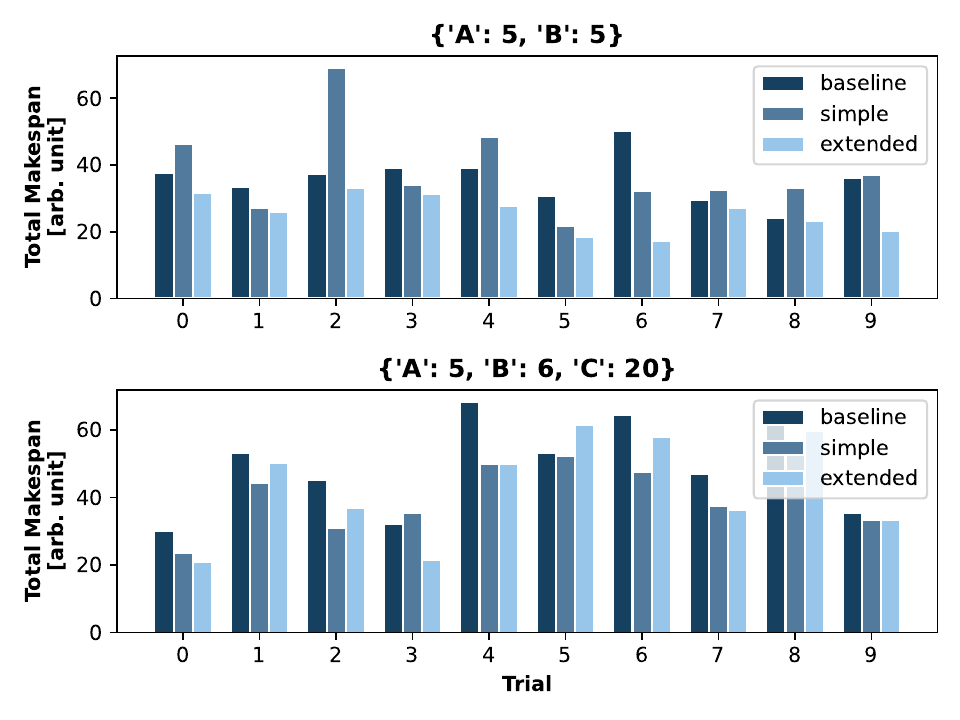}
            \caption{Results for real-valued $s_{jm}$ and $p_{jm}$.}
    \end{subfigure}
    \begin{subfigure}{0.45\textwidth}
        \includegraphics[scale=0.5]{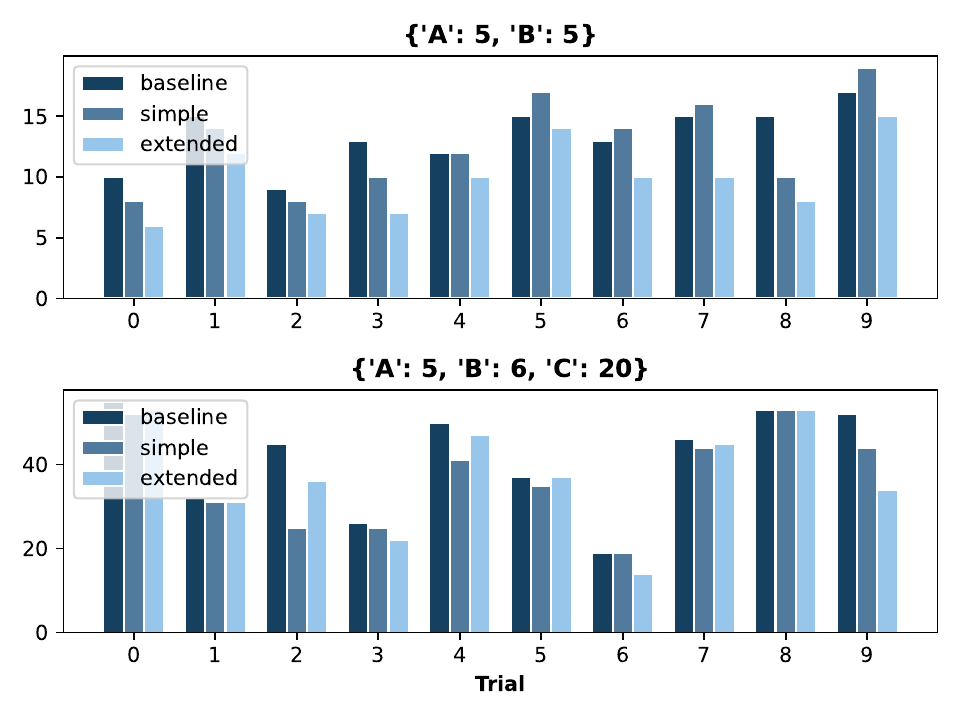}
        \caption{Results for integer-valued $s_{jm}$ and $p_{jm}$}
    \end{subfigure}
    % \subfigure{ % Results for real-valued $b_{jm}$ and $p_{jm}$.
    %     \includegraphics[scale=0.5]{figures/benchmark_results.pdf}
    %      %
    % }
    % \subfigure{ % Results for integer-valued $b_{jm}$ and $p_{jm}$
    %     \includegraphics[scale=0.5]{figures/benchmark_results_integer.pdf}
    % }
    \caption{Makespan results for the three configurations \emph{baseline}, \emph{simple} and \emph{extended} in two different settings. }
    \label{fig:benchmark_results}
\end{figure*}

\subsection{Simulation Results}

\cref{fig:benchmark_results} shows the results of the makespan optimization.
The full data is available in the code repository.

The \emph{extended} model consistently outperforms the \emph{simple} and \emph{baseline} algorithms.
The difference is more noticeable when we can parallelize multiple circuits.
In the setting with two \glspl{QPU}, circuits are similarly sized, and the makespan can be reduced by around \SI{25}{\percent} by the \emph{extended} model for real and integer inputs.
Due to the size imbalance, the second setting has potentially sequential circuits on the largest device, reducing the gain to \SI{12}{\percent} on average; for the real-valued input, the \emph{extended} model performs slightly better.
This also explains the outliers for the \emph{simple} model, where the setup times are potentially assumed to be worse for one device, which is then avoided entirely.
As a result, sequential execution decreases the performance, but a reduction of \SI{11}{\percent} on average can still be achieved.
In the worst-case setting (two devices with real-valued inputs), the baseline is faster than the \emph{simple} approach.

Besides usability, the time to solution is also an essential factor.
Due to the exponential overhead, the \emph{simple} model is roughly four magnitudes slower, and the \emph{extended} model is six magnitudes slower than the baseline.
Especially in the desired scenarios with considerable parallelization potential, the \emph{extended} model needs to evaluate multiple solutions.
A bottleneck is the configuration and invocation of Gurobi, which we call through a Python interface using the default parameters.
An approximation would likely suffice in a production environment, which we plan to implement in future iterations of \tool{}.

\section{Future Work} \label{sec:FutureWork}

The current state of \tool{} primarily serves as a proof of concept rather than an full-fledged implementation. 
There are several potential avenues for improvement. 
Initially, the scheduling functionality is limited to a batched or offline environment. 
Real systems with continuous job submissions regularly trigger rescheduling based on a previous schedule.
Such updates, as well as considering other criteria like preemption and priority, are necessary improvements.
Integrating circuit knitting with the scheduler, rather than keeping them separate components, could enhance efficiency, especially in optimizing cutting decisions for available hardware.
The exact solving of \gls{MILP} poses limitations when handling larger problems, suggesting the potential integration of heuristics for quicker solutions.

Expanding \tool{}'s scope beyond isolated \glspl{QPU} is a future goal, with plans for interprocess communication, exemplified by upcoming technologies like the IBM Flamingo chip~\cite{ibm2022}. 
To accommodate such advancements, substituting the knitting module with a distribution component becomes necessary. 
Additionally, as hardware sizes increase, \tool{} could schedule (partial) operations on error-corrected qubits instead of entire circuits.
However, it's important to note that within \tool{}'s current scope, there is no notion of logical circuits that cannot be cut or distributed over multiple devices.

\tool{}'s optimization ideally relies on accurate processing and setup time estimates. 
Unfortunately, such precise data is largely unavailable for most systems. 
A standardized interface for \glspl{QPU} would greatly aid in leveraging real-time data to influence optimal scheduling decisions.
This interface could also provide live fidelity information, which could be used during the mapping and cutting steps.

% \tool{} mainly serves as a proof of concept; it offers multiple directions of improvement.
% First, its scheduling only works in a batched environment.
% In a realistic environment, new jobs can be submitted anytime, which would need to trigger a rescheduling.
% Circuit knitting can also be integrated with the scheduler.
% Both components operate separately; the scheduler only receives circuits already resized to the devices.
% Cutting decisions can be optimized for the available hardware if both work together.
% The \gls{MILP} is also solved exactly; solving larger problems quickly becomes infeasible.
% In that case, heuristics can be employed to provide a useful solution quickly.
% Other metrics, such as individual qubit fidelities, can be implemented.

% Currently, \tool{} is designed for isolated \glspl{QPU}.
% In the future, interprocess communication will be possible, for example, in the planned Flamingo IBM chip~\cite{ibm2022}.
% \tool{} can be adapted by substituting the knitting module by a \emph{distribution} component.
% Similarly, with increasing hardware sizes, error correction becomes feasible.
% We do not consider logical circuits, which can not be cut or distributed over multiple devices.

% In the best case, \tool{} would operate based on accurate processing and setup times estimates.
% Unfortunately, this data is not available for most systems.
% While we can extrapolate from historical data, a standard interface for \glspl{QPU} is desirable.
% Based on this interface, real-time available data could influence the optimal schedule.

\section{Conclusion} \label{sec:Conclusion}
In this article, we present \tool{}, a software tool that can address multiple quantum hardware backends at the same time.
It automatically distributes batches of circuits over the available hardware, which are not necessarily the same type.
We formulate a \gls{MILP} to solve the scheduling problem, which arises when the circuits are resized to fit the hardware.
We prioritize minimizing the overall execution time and, therefore, emphasize hardware utilization.
In a set of benchmarks, we show an average makespan improvement of \SI{20}{\percent} compared to a baseline algorithm. 
\tool{} works as an end-to-end tool but can easily be integrated into existing infrastructures.

\section*{Acknowledgment}
The research is part of the Munich Quantum Valley~(MQV), which is supported by the Bavarian state government with funds from the Hightech Agenda Bayern Plus. Moreover, this project is also supported by the Federal Ministry for Economic Affairs and Climate Action on the basis of a decision by the German Bundestag through project QuaST, as well as by the Bavarian Ministry of Economic Affairs, Regional Development and Energy with funds from the Hightech Agenda Bayern. Further funding comes from the German Federal Ministry of Education and Research~(BMBF) through the MUNIQC-SC project.

% ++++++++++++++++++++++++++++++++++++++++++++++++++++++++++++++++++++++++++++
% ++++++++++++++++++++++++++++++++++++++++++++++++++++++++++++++++++++++++++++
% ++++++++++++++++++++++++++++++++++++++++++++++++++++++++++++++++++++++++++++
% \section*{References}
% \vspace{1.6cm}
% \IEEEtriggeratref{9}
\bibliographystyle{IEEEtran}
\bibliography{references}
\end{document}

%% file: glosfile.tex
\newacronym{MILP}{MILP}{mixed-integer linear program}
\newacronym{AWG}{AWG}{arbitrary waveform generator}
\newacronym{NISQ}{NISQ}{noisy intermediate-scale quantum}

\newacronym{CPU}{CPU}{central processing unit}
\newacronym{GPU}{GPU}{graphics processing unit}
%\newacronym{VMemory}{VMemory}{video memory}
\newacronym{QPU}{QPU}{quantum processing unit}
%\newacronym{NIC}{NIC}{network interface controller}
\newacronym{HPC}{HPC}{high performance computing}
%\newacronym{GPGPU}{GPGPU}{general purpose graphics processing Unit}
\newacronym{LRZ}{LRZ}{Leibniz Supercomputing Centre}
\newacronym{DSL}{DSL}{domain-specific language}
\newacronym{eDSL}{eDSL}{embedded domain-specific language}
\newacronym{MPI}{MPI}{Message Passing Interface}
\newacronym{QRAM}{QRAM}{quantum random access machine}
\newacronym{HQCC}{HQCC}{heterogeneous quantum-classical computation}
\newacronym{API}{API}{application programming interface}
%\newacronym{QPL}{QPL}{quantum programming language}
\newacronym{QPT}{QPT}{quantum programming tool}
\newacronym{QC}{QC}{quantum computing}
%\newacronym{VQE}{VQE}{variational quantum eigensolver}
\newacronym{ISA}{ISA}{instruction set architecture}
%\newacronym{OpenQASM}{OpenQASM}{Open Quantum Assembly Language}
%\newacronym{CISC}{CISC}{complex instruction set computer}
\newacronym{RISC}{RISC}{reduced instruction set computer}
% \newacronym{ARM}{ARM}{advanced RISC machine}
\newacronym{RAM}{RAM}{random access memory}
%\newacronym{AQC}{AQC}{Adiabatic Quantum Computing}
\newacronym{IR}{IR}{intermediate representation}
\newacronym{SQRAM}{SQRAM}{sequential QRAM}
\newacronym{HPCQC}{HPCQC}{high performance computing - quantum computing}
\newacronym{QIR}{QIR}{Quantum Intermediate Representation}
\newacronym{SQIR}{SQIR}{standardized quantum intermediate representation}
\newacronym{AOT}{AOT}{ahead-of-time}
\newacronym{JIT}{JIT}{just-in-time}

%% file: main.bbl
\begin{thebibliography}{10}
\providecommand{\url}[1]{#1}
\csname url@samestyle\endcsname
\providecommand{\newblock}{\relax}
\providecommand{\bibinfo}[2]{#2}
\providecommand{\BIBentrySTDinterwordspacing}{\spaceskip=0pt\relax}
\providecommand{\BIBentryALTinterwordstretchfactor}{4}
\providecommand{\BIBentryALTinterwordspacing}{\spaceskip=\fontdimen2\font plus
\BIBentryALTinterwordstretchfactor\fontdimen3\font minus
  \fontdimen4\font\relax}
\providecommand{\BIBforeignlanguage}[2]{{%
\expandafter\ifx\csname l@#1\endcsname\relax
\typeout{** WARNING: IEEEtran.bst: No hyphenation pattern has been}%
\typeout{** loaded for the language `#1'. Using the pattern for}%
\typeout{** the default language instead.}%
\else
\language=\csname l@#1\endcsname
\fi
#2}}
\providecommand{\BIBdecl}{\relax}
\BIBdecl

\bibitem{shor1994algorithms}
\BIBentryALTinterwordspacing
P.~W. Shor, ``Algorithms for quantum computation: discrete logarithms and
  factoring,'' in \emph{Proceedings 35th annual symposium on foundations of
  computer science}.\hskip 1em plus 0.5em minus 0.4em\relax Santa Fe, NM, USA:
  IEEE, 1994, pp. 124--134. [Online]. Available:
  \url{https://doi.org/10.1109/SFCS.1994.365700}
\BIBentrySTDinterwordspacing

\bibitem{fowler2012}
\BIBentryALTinterwordspacing
A.~G. Fowler, M.~Mariantoni, J.~M. Martinis, and A.~N. Cleland, ``Surface
  codes: Towards practical large-scale quantum computation,'' \emph{Phys. Rev.
  A}, vol.~86, p. 032324, Sep 2012. [Online]. Available:
  \url{https://link.aps.org/doi/10.1103/PhysRevA.86.032324}
\BIBentrySTDinterwordspacing

\bibitem{kjaergaard2020}
\BIBentryALTinterwordspacing
M.~Kjaergaard, M.~E. Schwartz, J.~Braum\"{u}ller, P.~Krantz, J.~I.-J. Wang,
  S.~Gustavsson, and W.~D. Oliver, ``Superconducting qubits: Current state of
  play,'' \emph{Annual Review of Condensed Matter Physics}, vol.~11, no.~1, pp.
  369--395, 2020. [Online]. Available:
  \url{https://doi.org/10.1146/annurev-conmatphys-031119-050605}
\BIBentrySTDinterwordspacing

\bibitem{Wu2021}
\BIBentryALTinterwordspacing
X.~Wu, X.~Liang, Y.~Tian, F.~Yang, C.~Chen, Y.-C. Liu, M.~K. Tey, and L.~You,
  ``A concise review of rydberg atom based quantum computation and quantum
  simulation*,'' \emph{Chinese Physics B}, vol.~30, no.~2, p. 020305, feb 2021.
  [Online]. Available: \url{https://dx.doi.org/10.1088/1674-1056/abd76f}
\BIBentrySTDinterwordspacing

\bibitem{Bruzewicz2019}
\BIBentryALTinterwordspacing
C.~D. Bruzewicz, J.~Chiaverini, R.~McConnell, and J.~M. Sage, ``{Trapped-ion
  quantum computing: Progress and challenges},'' \emph{Applied Physics
  Reviews}, vol.~6, no.~2, p. 021314, 05 2019. [Online]. Available:
  \url{https://doi.org/10.1063/1.5088164}
\BIBentrySTDinterwordspacing

\bibitem{bandic2023}
\BIBentryALTinterwordspacing
M.~Bandic, L.~Prielinger, J.~Nüßlein, A.~Ovide, S.~Rodrigo, S.~Abadal, H.~van
  Someren, G.~Vardoyan, E.~Alarcon, C.~G. Almudever, and S.~Feld, ``Mapping
  quantum circuits to modular architectures with qubo,'' 2023, unpublished.
  [Online]. Available: \url{https://doi.org/10.48550/arXiv.2305.06687}
\BIBentrySTDinterwordspacing

\bibitem{bhoumik2023}
\BIBentryALTinterwordspacing
D.~Bhoumik, R.~Majumdar, A.~Saha, and S.~Sur-Kolay, ``Distributed scheduling of
  quantum circuits with noise and time optimization,'' 2023, unpublished.
  [Online]. Available: \url{https://doi.org/10.48550/arXiv.2309.06005}
\BIBentrySTDinterwordspacing

\bibitem{graham1979}
\BIBentryALTinterwordspacing
R.~Graham, E.~Lawler, J.~Lenstra, and A.~Kan, ``Optimization and approximation
  in deterministic sequencing and scheduling: a survey,'' in \emph{Discrete
  Optimization II}, ser. Annals of Discrete Mathematics, P.~Hammer, E.~Johnson,
  and B.~Korte, Eds.\hskip 1em plus 0.5em minus 0.4em\relax Elsevier, 1979,
  vol.~5, pp. 287--326. [Online]. Available:
  \url{https://www.sciencedirect.com/science/article/pii/S016750600870356X}
\BIBentrySTDinterwordspacing

\bibitem{fanjul2011}
\BIBentryALTinterwordspacing
L.~Fanjul-Peyro and R.~Ruiz, ``Size-reduction heuristics for the unrelated
  parallel machines scheduling problem,'' \emph{Comput. Oper. Res.}, vol.~38,
  no.~1, p. 301–309, jan 2011. [Online]. Available:
  \url{https://doi.org/10.1016/j.cor.2010.05.005}
\BIBentrySTDinterwordspacing

\bibitem{rodriguez2013}
\BIBentryALTinterwordspacing
F.~J. Rodriguez, M.~Lozano, C.~Blum, and C.~Garc\'{\i}a-Mart\'{\i}nez, ``An
  iterated greedy algorithm for the large-scale unrelated parallel machines
  scheduling problem,'' \emph{Comput. Oper. Res.}, vol.~40, no.~7, p.
  1829–1841, jul 2013. [Online]. Available:
  \url{https://doi.org/10.1016/j.cor.2013.01.018}
\BIBentrySTDinterwordspacing

\bibitem{alharkan.2019}
\BIBentryALTinterwordspacing
I.~M. Al-harkan and A.~A. Qamhan, ``Optimize unrelated parallel machines
  scheduling problems with multiple limited additional resources,
  sequence-dependent setup times and release date constraints,'' \emph{IEEE
  Access}, vol.~7, pp. 171\,533--171\,547, 2019. [Online]. Available:
  \url{https://doi.org/10.1109/ACCESS.2019.2955975}
\BIBentrySTDinterwordspacing

\bibitem{Lenstra1990}
\BIBentryALTinterwordspacing
J.~K. Lenstra, D.~B. Shmoys, and {\'E}.~Tardos, ``Approximation algorithms for
  scheduling unrelated parallel machines,'' \emph{Mathematical Programming},
  vol.~46, no.~1, pp. 259--271, Jan 1990. [Online]. Available:
  \url{https://doi.org/10.1007/BF01585745}
\BIBentrySTDinterwordspacing

\bibitem{glass1994}
\BIBentryALTinterwordspacing
C.~Glass, C.~Potts, and P.~Shade, ``Unrelated parallel machine scheduling using
  local search,'' \emph{Mathematical and Computer Modelling}, vol.~20, no.~2,
  pp. 41--52, 1994. [Online]. Available:
  \url{https://www.sciencedirect.com/science/article/pii/0895717794902054}
\BIBentrySTDinterwordspacing

\bibitem{Yoo2003}
\BIBentryALTinterwordspacing
A.~B. Yoo, M.~A. Jette, and M.~Grondona, ``Slurm: Simple linux utility for
  resource management,'' in \emph{Job Scheduling Strategies for Parallel
  Processing}, D.~Feitelson, L.~Rudolph, and U.~Schwiegelshohn, Eds.\hskip 1em
  plus 0.5em minus 0.4em\relax Berlin, Heidelberg: Springer Berlin Heidelberg,
  2003, pp. 44--60. [Online]. Available:
  \url{https://doi.org/10.1007/10968987_3}
\BIBentrySTDinterwordspacing

\bibitem{humble2021}
\BIBentryALTinterwordspacing
T.~S. Humble, A.~McCaskey, D.~I. Lyakh, M.~Gowrishankar, A.~Frisch, and
  T.~Monz, ``Quantum computers for high-performance computing,'' \emph{IEEE
  Micro}, vol.~41, no.~5, p. 15–23, sep 2021. [Online]. Available:
  \url{https://doi.org/10.1109/MM.2021.3099140}
\BIBentrySTDinterwordspacing

\bibitem{QIRSpec2021}
\BIBentryALTinterwordspacing
{QIR Alliance}, \emph{{QIR Specification}}, 2021, also see
  \url{https://qir-alliance.org}. [Online]. Available:
  \url{https://github.com/qir-alliance/qir-spec}
\BIBentrySTDinterwordspacing

\bibitem{peng2020}
\BIBentryALTinterwordspacing
T.~Peng, A.~W. Harrow, M.~Ozols, and X.~Wu, ``Simulating large quantum circuits
  on a small quantum computer,'' \emph{Phys. Rev. Lett.}, vol. 125, p. 150504,
  Oct 2020. [Online]. Available:
  \url{https://link.aps.org/doi/10.1103/PhysRevLett.125.150504}
\BIBentrySTDinterwordspacing

\bibitem{mitarai2021}
\BIBentryALTinterwordspacing
K.~Mitarai and K.~Fujii, ``Constructing a virtual two-qubit gate by sampling
  single-qubit operations,'' \emph{New Journal of Physics}, vol.~23, no.~2, p.
  023021, feb 2021. [Online]. Available:
  \url{https://dx.doi.org/10.1088/1367-2630/abd7bc}
\BIBentrySTDinterwordspacing

\bibitem{Ufrecht2023}
\BIBentryALTinterwordspacing
C.~Ufrecht, M.~Periyasamy, S.~Rietsch, D.~D. Scherer, A.~Plinge, and
  C.~Mutschler, ``Cutting multi-control quantum gates with {ZX} calculus,''
  \emph{{Quantum}}, vol.~7, p. 1147, Oct. 2023. [Online]. Available:
  \url{https://doi.org/10.22331/q-2023-10-23-1147}
\BIBentrySTDinterwordspacing

\bibitem{harada2023optimal}
\BIBentryALTinterwordspacing
H.~Harada, K.~Wada, and N.~Yamamoto, ``Optimal parallel wire cutting without
  ancilla qubits,'' 2023, unpublished. [Online]. Available:
  \url{https://doi.org/10.48550/arXiv.2303.07340}
\BIBentrySTDinterwordspacing

\bibitem{Qiskit}
\BIBentryALTinterwordspacing
{Qiskit contributors}, ``Qiskit: An open-source framework for quantum
  computing,'' 2023. [Online]. Available:
  \url{https://zenodo.org/doi/10.5281/zenodo.2573505}
\BIBentrySTDinterwordspacing

\bibitem{circuit-knitting-toolbox}
\BIBentryALTinterwordspacing
L.~Bello, A.~M. Bra\'{n}czyk, S.~Bravyi, A.~{Carrera Vazquez}, A.~Eddins, D.~J.
  Egger, B.~Fuller, J.~Gacon, J.~R. Garrison, J.~R. Glick, T.~P. Gujarati,
  I.~Hamamura, A.~I. Hasan, T.~Imamichi, C.~Johnson, I.~Liepuoniute,
  O.~Lockwood, M.~Motta, C.~D. Pemmaraju, P.~Rivero, M.~Rossmannek, T.~L.
  Scholten, S.~Seelam, I.~Sitdikov, D.~Subramanian, W.~Tang, and S.~Woerner,
  ``{Circuit Knitting Toolbox},'' 2023. [Online]. Available:
  \url{https://zenodo.org/doi/10.5281/zenodo.7987996}
\BIBentrySTDinterwordspacing

\bibitem{gurobi}
\BIBentryALTinterwordspacing
{Gurobi Optimization, LLC}, ``{Gurobi Optimizer Reference Manual},'' 2023.
  [Online]. Available: \url{https://www.gurobi.com}
\BIBentrySTDinterwordspacing

\bibitem{Barredo2018}
\BIBentryALTinterwordspacing
D.~Barredo, V.~Lienhard, S.~de~L{\'e}s{\'e}leuc, T.~Lahaye, and A.~Browaeys,
  ``Synthetic three-dimensional atomic structures assembled atom by atom,''
  \emph{Nature}, vol. 561, no. 7721, pp. 79--82, Sep 2018. [Online]. Available:
  \url{https://doi.org/10.1038/s41586-018-0450-2}
\BIBentrySTDinterwordspacing

\bibitem{quetschlich2023mqtbench}
\BIBentryALTinterwordspacing
N.~Quetschlich, L.~Burgholzer, and R.~Wille, ``{MQT} {B}ench: {B}enchmarking
  {S}oftware and {D}esign {A}utomation {T}ools for {Q}uantum {C}omputing,''
  \emph{{Quantum}}, vol.~7, p. 1062, Jul. 2023. [Online]. Available:
  \url{https://doi.org/10.22331/q-2023-07-20-1062}
\BIBentrySTDinterwordspacing

\bibitem{baker1985}
\BIBentryALTinterwordspacing
B.~S. Baker, ``A new proof for the first-fit decreasing bin-packing
  algorithm,'' \emph{Journal of Algorithms}, vol.~6, no.~1, pp. 49--70, 1985.
  [Online]. Available:
  \url{https://www.sciencedirect.com/science/article/pii/0196677485900185}
\BIBentrySTDinterwordspacing

\bibitem{ibm2022}
\BIBentryALTinterwordspacing
J.~Gambetta. (2022) Expanding the ibm quantum roadmap to anticipate the future
  of quantum-centric supercomputing. [Online]. Available:
  \url{https://research.ibm.com/blog/ibm-quantum-roadmap-2025}
\BIBentrySTDinterwordspacing

\end{thebibliography}
